\pdfoutput=1
\documentclass{JINST}
\let\ifpdf\relax

\usepackage{commath}

\newcommand{\farcs}{.\!\!^{\prime\prime}}

\title{Impact of chromatic effects on galaxy shape measurements}

\author{J. E. Meyers, P. R. Burchat\\
Kavli Institute for Particle Astrophysics and Cosmology, Department of Physics, Stanford University, Stanford CA 94305\\
E-mail: \email{jmeyers3@stanford.edu}}

\abstract{Current and future imaging surveys will measure cosmic shear with a statistical precision that demands a deeper understanding of potential systematic biases in galaxy shape measurements than has been achieved to date.
We investigate the effects of using the point spread function (PSF) measured with stars to determine the shape of a galaxy that has a different spectral energy distribution (SED) than the star.
We demonstrate that a wavelength dependent PSF size, for example as may originate from atmospheric seeing or the diffraction limit of the primary aperture, can introduce significant shape measurement biases.
This analysis shows that even small wavelength dependencies in the PSF may introduce biases, and hence that achieving the ultimate precision for weak lensing from current and future imaging surveys will require a detailed understanding of the wavelength dependence of the PSF from {\em all} sources, including the CCD sensors. }

\keywords{Cosmology; Weak Gravitational Lensing; Cosmic Shear; Spectral Energy Distributions; Point Spread Functions}

\begin{document}

\section{Introduction}\label{sec:intro}

A goal of large astronomical imaging surveys is to constrain cosmological parameters by measuring the small departure from statistical isotropy of the shapes and orientations of distant galaxies, induced by the gravitational lensing from foreground large-scale structure.
The shapes of galaxy images, however, are not only affected by cosmic shear (typically a $\lesssim 1\%$ shift in the major-to-minor axis ratio), but are also determined by the combined point spread function (PSF) due to the atmosphere (for ground-based instruments), telescope optics, and the image sensor -- together often a few $\%$ shift.
The size and shape of this additional convolution kernel is typically determined from the observed images of stars, which are effectively point sources before being smeared by the PSF.
Galaxy images can then be deconvolved with the estimated convolution kernel.
Implicit in this approach is the assumption that the kernel for galaxies is the same as the kernel for stars.
If the PSF is dependent on wavelength, this assumption is violated since stars and galaxies have different spectral energy distributions (SEDs) and hence different PSFs.
Correcting observed images with the incorrect PSF can lead to biases in shape measurements.
To predict and eliminate these biases, we must understand all PSF contributions that depend on wavelength.

In this paper, we illustrate the biases that can arise from a wavelength-dependent PSF by considering a particular class of chromatic effects.
In Section \ref{sec:size}, we describe our PSF model, and in Section \ref{sec:biases} estimate the resulting galaxy shape biases.
In Section \ref{sec:requirements}, we investigate how these biases affect current and proposed weak lensing surveys.
We conclude in Section \ref{sec:conclusion}.

\section{PSF-size -- wavelength relation}\label{sec:size}

One possible chromatic effect is a dependence of PSF size (i.e., full-width-half-maximum (FWHM)) on wavelength.
For example, the Kolmogorov theory of atmospheric turbulence predicts that atmospheric seeing should scale like
\begin{equation}
\mathrm{FWHM}_\mathrm{seeing} \propto \lambda^{-1/5},
\end{equation}
while a diffraction-limited telescope will have a chromatic PSF that scales like
\begin{equation}
    \mathrm{FWHM}_\mathrm{diffraction\, limit} \propto \lambda^{+1}.
\end{equation}
The power law indices for complete systems, including contributions from atmosphere (for ground-based telescopes), optics, and sensors may be somewhat different.
For example, \cite{Voigt++12, Cypriano++10} find that the FWHM of the Euclid space telescope \cite{Laureijs++11} PSF varies approximately like $\lambda^{+0.6}$.
Detailed measurements of chromatic effects in CCDs, which naturally arise from the wavelength dependence of the silicon absorption length, are actively being pursued.

To keep our analysis generic, we investigate the impact of a PSF with a power-law wavelength dependence:
\begin{equation}
  \mathrm{FWHM} \propto \lambda^\alpha.
\end{equation}

\section{Shape measurement biases}\label{sec:biases}

Weak gravitational lensing is frequently analyzed through its effect on combinations of the second central moments $I_{\mu\nu}$ of a galaxy's surface brightness distribution given by
\begin{equation}
  \label{eq:M2}
  I_{\mu \nu} = \frac{1}{f}\int{\dif{x} \dif{y} I(x,y)(\mu - \bar{\mu})(\nu - \bar{\nu})},
\end{equation}
where $\mu$ and $\nu$ each refer to $x$ or $y$.
The centroids $\bar{\mu}$ and $\bar{\nu}$ and the total flux $f$ of the surface brightness distribution are given by
\begin{equation}
  \label{eq:M1}
  \bar{\mu} = \frac{1}{f}\int{\dif{x} \dif{y} I(x,y)\mu},
\end{equation}
\begin{equation}
  \label{eq:M0}
  f = \int{\dif{x} \dif{y} I(x,y)}.
\end{equation}

Two important combinations of second central moments are the second-moment square radius $r^2$ and the complex ellipticity $\boldsymbol{\epsilon} = \epsilon_1 + \mathrm{i}\epsilon_2$:
\begin{equation}
  \label{eq:r2}
  r^2 = I_{xx} + I_{yy},
\end{equation}
\begin{equation}
  \label{eq:e1}
  \epsilon_1 = \frac{I_{xx} - I_{yy}}{I_{xx} + I_{yy}},
\end{equation}
\begin{equation}
  \label{eq:e2}
  \epsilon_2 = \frac{2 I_{xy}}{I_{xx} + I_{yy}}.
\end{equation}
With this definition of ellipticity, an object with perfectly elliptical isophotes and ratio $q$ of minor to major axes ($0 \le q \le 1$) will have ellipticity magnitude equal to
\begin{equation}
  \label{eq:q_to_ellip}
  |\boldsymbol{\epsilon}| = \frac{1-q^2}{1+q^2}.
\end{equation}
A galaxy's apparent (lensed) ellipticity $\boldsymbol{\epsilon^{(a)}}$ is related to its intrinsic (unlensed) ellipticity $\boldsymbol{\epsilon^{(i)}}$ in the presence of gravitational lensing shear $\boldsymbol{\gamma} = \gamma_1 + \mathrm{i} \gamma_2$ and convergence $\kappa$ via
\begin{equation}
  \boldsymbol{\epsilon^{(a)}} = \frac{\boldsymbol{\epsilon^{(i)}} - 2 \boldsymbol{g} + \boldsymbol{g}^2 \boldsymbol{\epsilon^{(i)}}^*}{1+|g|^2-2 \Re(\boldsymbol{g}\boldsymbol{\epsilon^{(i)}}^*)}
\end{equation}
where $\boldsymbol{g} = \boldsymbol{\gamma}/(1-\kappa)$ is the reduced shear\cite{Schneider+Seitz95}.
Under the assumption that intrinsic galaxy ellipticities are isotropically distributed, the reduced shear is related to the mean of the sheared ellipticities by $\left\langle \boldsymbol{\epsilon^{(a)}} \right\rangle \approx 2 \boldsymbol{\gamma}$.
The correlation function or power spectrum of shears forms the cosmologically pertinent statistic.

Since we are interested in measuring the ellipticity of the surface brightness distribution of a lensed galaxies before convolution with the PSF, but only have access to the surface brightness distribution after convolution, we must apply a PSF correction.
This can be accomplished by applying $I^\mathrm{gal}_{\mu\nu} = I^\mathrm{obs}_{\mu\nu} - I^\mathrm{PSF}_{\mu\nu}$, which holds exactly for unweighted second moments and where the second moments of the PSF can be estimated from observations of stars.
In practice, noisy data require one to use weighted second moments, rendering the above relation only approximate, but still useful (and in fact still exact in the case of Gaussian profiles and weight functions).
Small differences between stellar and galactic SEDs will induce small systematic errors into estimates of the PSF size $\delta r^2_\mathrm{PSF}$, however.
These size errors then propagate into shape errors \cite{Paulin-Henriksson++08} as
\begin{equation}
    \label{eq:PH08}
    \delta \epsilon_\mathrm{sys} = (\epsilon_\mathrm{gal}-\epsilon_\mathrm{PSF}) \left(\frac{\delta r^2_\mathrm{PSF}}{r^2_\mathrm{gal}}\right).
\end{equation}
We parameterize the bias in the shear in terms of multiplicative and additive terms, $\hat{\gamma}_i=\gamma_i(1+m_i)+c_i$,
$i=1,2$, where $\hat{\boldsymbol{\gamma}}$ is the estimator for the true shear $\boldsymbol{\gamma}$.
The shear bias due to a misestimated PSF size can then be written
\begin{equation}
  \label{eq:m1m2}
  m_1=m_2=\frac{\delta r^2_\mathrm{PSF}}{r^2_\mathrm{PSF}}\frac{r^2_\mathrm{PSF}}{r^2_\mathrm{gal}},
\end{equation}
\begin{equation}
  \label{eq:ci}
  c_i=-\frac{\epsilon^\mathrm{PSF}_i}{2}\frac{\delta r^2_\mathrm{PSF}}{r^2_\mathrm{PSF}}\frac{r^2_\mathrm{PSF}}{ r^2_\mathrm{gal}},
\end{equation}
where we have assumed that $\hat\gamma_1$ ($\hat\gamma_2$) is independent of $\gamma_2$ ($\gamma_1$).
Note that we have multiplied and divided by $r^2_\mathrm{PSF}$ in both of these expressions in order to isolate a term, $\delta r^2_\mathrm{PSF}/r^2_\mathrm{PSF}$, which is independent of the conditions (i.e. the fixed-wavelength seeing) of a particular observation.
In other words, this term depends on the SEDs of the detected photons of the star and galaxy in question, and on the power law index of the seeing--wavelength relation, but not on the absolute size of the PSF.

\begin{figure}[tbp]
  \centering
  \includegraphics[width=.8\textwidth]{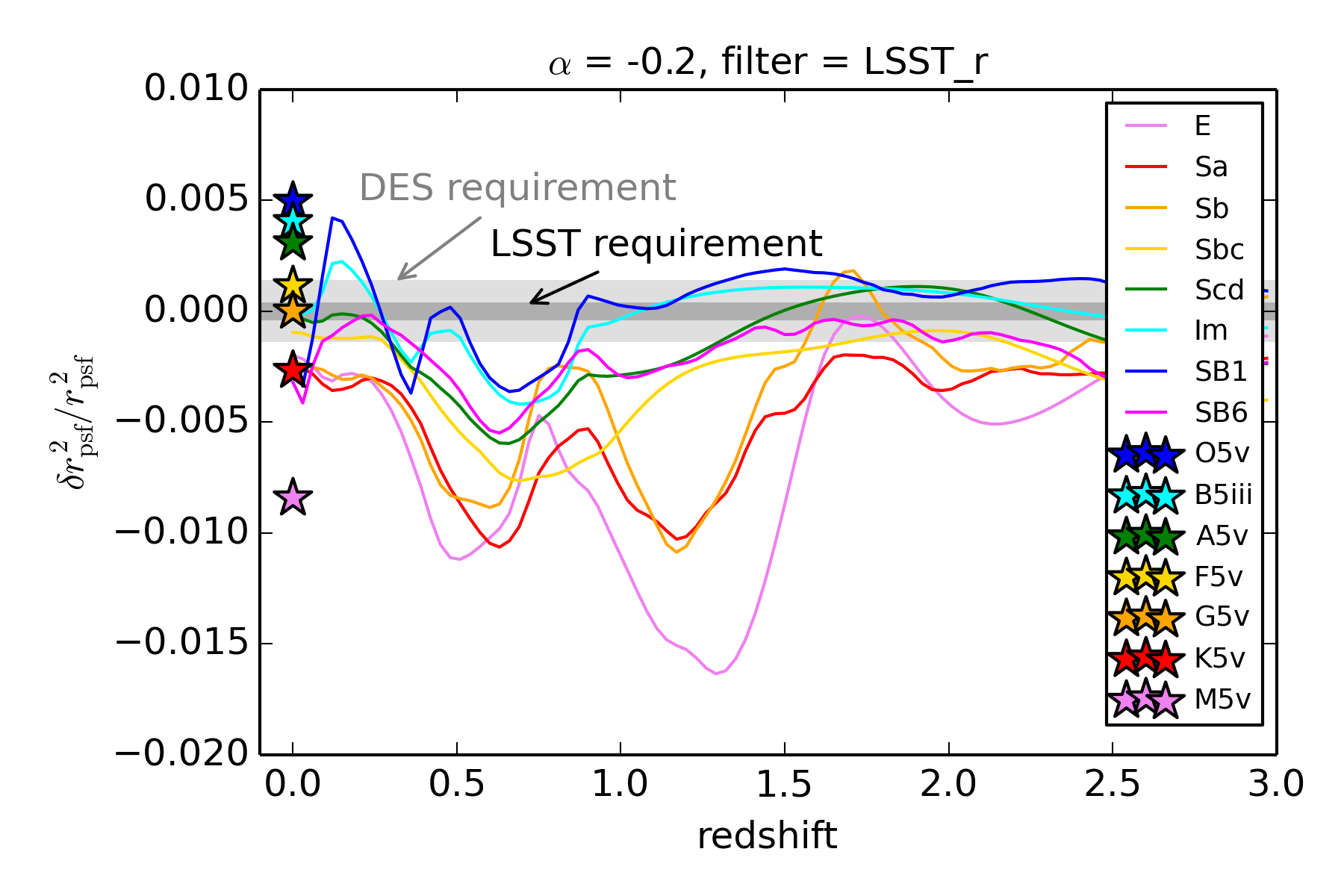}
  \caption{
    Fractional shifts in $r^2_\mathrm{PSF}$ due to chromatic seeing, calculated for the LSST $r$-band (which is very similar to the DES $r$-band).
    Shifts are (arbitrarily) normalized relative to a G5v star.
    Star symbols at redshift 0 represent stellar SEDs from \cite{Pickles98}.
    Lines represent galactic SEDs from \cite{Coleman++80} and \cite{Kinney++96}.
    The requirements for DES and LSST are overplotted.
    Note that only the widths of the requirements relative to the scatter of the stellar and galactic PSFs are relevant, and not their absolute positions along the y-axis.
    In the $i$ band, which is the only other band planned for shape measurement with LSST, the magnitudes of the shifts are about 50\% smaller, but the requirement is the same.}
  \label{fig:f1}
\end{figure}

\begin{figure}[tbp]
  \centering
  \includegraphics[width=.8\textwidth]{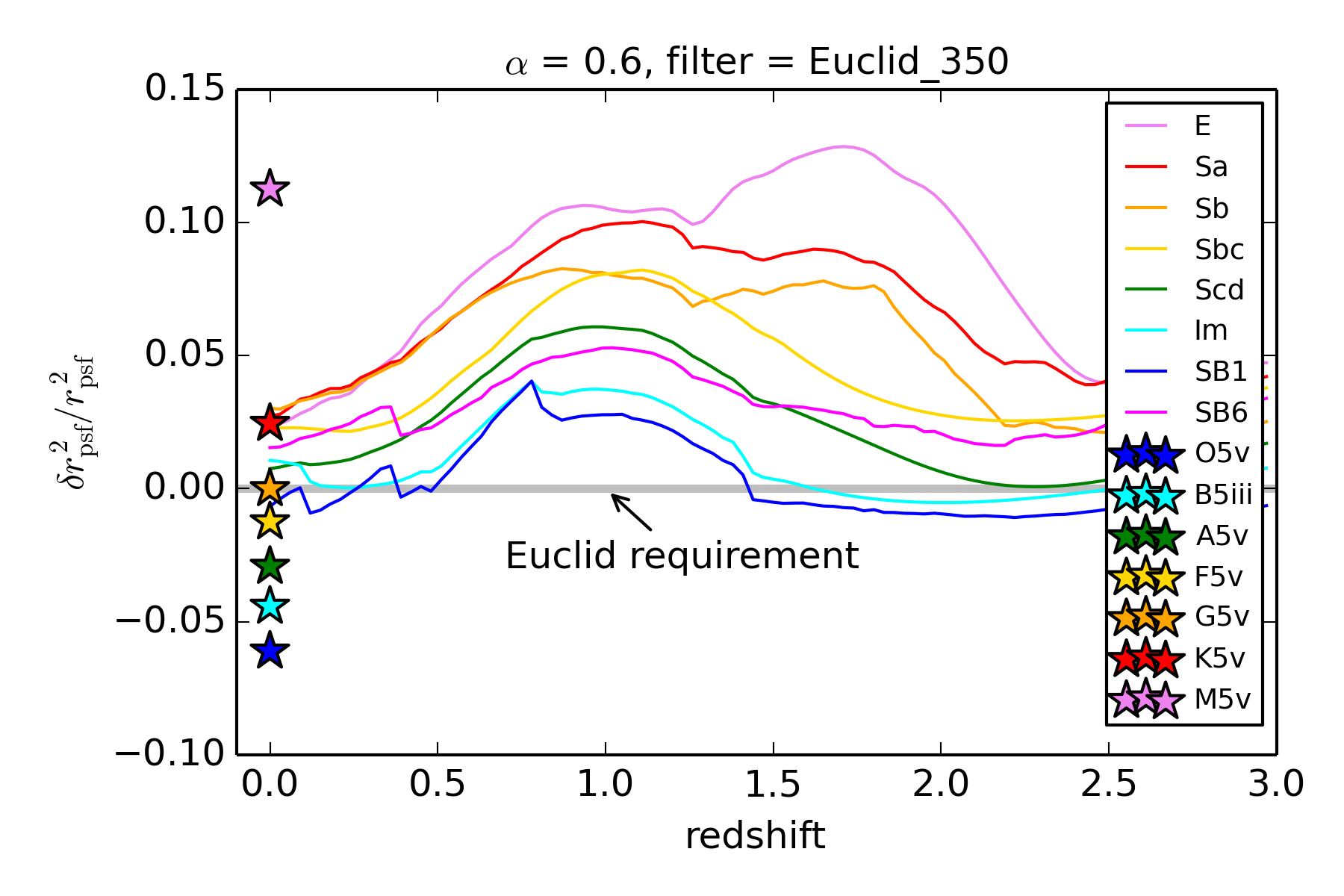}
  \caption{
    Fractional shifts in $\delta r^2_\mathrm{PSF}$ due to a Euclid-like PSF with $\mathrm{FWHM} \propto \lambda^{+0.6}$, calculated for a Euclid-like 350nm-wide optical band (simulated as a tophat function with throughput between 550nm and 900nm).
    Shifts are (arbitrarily) normalized relative to a G5v star.
    Symbols, lines, and SEDs are the same as in Figure 1.
    The requirement for Euclid is overplotted.
    Note that only the width of the requirement relative to the scatter of the stellar and galactic PSFs is relevant, and not its absolute position along the y-axis.
    The wide optical band is the only band planned for shape measurement with Euclid.}
  \label{fig:f2}
\end{figure}

The second moments of the PSF for a given SED are the photon-weighted sums of second moments at each wavelength, which we assume are rescalings of the PSF at a reference wavelength $\lambda_0$:
\begin{equation}
    I_{\mu\nu}^\mathrm{PSF} = I_{\mu\nu}^\mathrm{PSF, \lambda_0} \frac{\int p(\lambda) (\lambda/\lambda_0)^{2 \alpha}\dif{\lambda}}{\int p(\lambda)\dif{\lambda}},
\end{equation}
or analogously for the second-moment square radius,
\begin{equation}
    \label{eq:r2PSF}
    r^2_\mathrm{PSF} = r^2_\mathrm{PSF, \lambda_0} \frac{\int p(\lambda) (\lambda/\lambda_0)^{2 \alpha}\dif{\lambda}}{\int p(\lambda)\dif{\lambda}},
\end{equation}
In these expressions, $p(\lambda)$ is the wavelength distribution of detected photons, i.e. the source photons multiplied by the system throughput.
In Figure \ref{fig:f1} and \ref{fig:f2} we use Equation \ref{eq:r2PSF} to compare the sizes of PSFs for representative stellar and galactic SEDs, both for a ground-based experiment with chromatic seeing ($\alpha=-0.2$), and a Euclid-like experiment whose PSF includes a chromatic contribution from the primary aperture diffraction limit ($\alpha=+0.6$).

\section{Survey requirements}\label{sec:requirements}

The sensitivity of a given survey to systematic shape biases depends on its statistical power, which depends primarily on the survey's area, depth, and effective number density of galaxies ($n_\mathrm{eff}$).
In Table \ref{tab:surveys} we use the formulae from \cite{Amara+Refregier08} to estimate the tolerable multiplicative and additive biases of a few current and future weak lensing surveys, given their area (in square degrees), median redshift $z_m$, and $n_\mathrm{eff}$ (number per square arcmin) \cite{DES++11, Chang++13, Laureijs++11}.
The requirements are set such that the systematic uncertainties on measurements of the dark energy equation of state parameter $w$ are equal to the statistical uncertainties.

\begin{table}[tbp]
    \centering
    \caption{Survey descriptions and shear bias tolerances.}
    \smallskip
    \begin{tabular}{|lccccc|}
        \hline
        Survey & Area  & $n_\mathrm{eff}$ & $z_m$ & $m_i$ & $c_i$  \\
        \hline
        DES    & 5000  & 12               & 0.7   & 0.004 & 0.0006 \\
        LSST   & 18000 & 30               & 0.9   & 0.001 & 0.0003 \\
        Euclid & 15000 & 30               & 0.9   & 0.001 & 0.0003 \\
        \hline
    \end{tabular}
    \label{tab:surveys}
\end{table}

\begin{table}[tbp]
    \centering
    \caption{Typical survey PSF and galaxy sizes, and PSF size misestimate tolerances.}
    \smallskip
    \begin{tabular}{|lccc|}
        \hline
        Survey & $r^2_\mathrm{PSF}$ & $r^2_\mathrm{gal}$ & $\delta r^2_\mathrm{PSF}/r^2_\mathrm{PSF}$ requirement \\
        \hline
        DES    & $(0\farcs8)^2$     & $(0\farcs47)^2$    & 0.0014 \\
        LSST   & $(0\farcs7)^2$     & $(0\farcs39)^2$    & 0.0004 \\
        Euclid & $(0\farcs2)^2$     & $(0\farcs23)^2$    & 0.0016 \\
        \hline
    \end{tabular}
    \label{tab:requirements}
\end{table}

The final ingredient needed to set a requirement on $\delta r^2_\mathrm{PSF}/r^2_\mathrm{PSF}$ is the ratio of the typical survey PSF size to the typical galaxy size, $r^2_\mathrm{PSF}/r^2_\mathrm{gal}$.
Generically, we can assume that this ratio is order unity, as surveys will naturally attempt to measure the shapes of galaxies down to their resolution limit.
For the multiplicative bias, this implies a requirement on $\delta r^2_\mathrm{PSF}/r^2_\mathrm{PSF}$ approximately equal to the requirement on $m$.
Despite the fact that survey additive bias requirements are numerically smaller than multiplicative bias requirements in Table \ref{tab:surveys}, requirements on $\delta r^2_\mathrm{PSF}/r^2_\mathrm{PSF}$ coming from additive bias constraints are generally more forgiving, as the extra factor of $\epsilon^\mathrm{PSF}/2$ in Equation \ref{eq:ci} compared to Equation \ref{eq:m1m2} is usually small.
In Table \ref{tab:requirements}, we estimate $r^2_\mathrm{PSF}$, $r^2_\mathrm{gal}$, and the resulting requirement on $\delta r^2_\mathrm{PSF}/r^2_\mathrm{PSF}$.
Estimates of $r^2_\mathrm{PSF}$ are taken from \cite{DES++11}, \cite{Chang++13}, and \cite{Laureijs++11}.
Estimates of $r^2_\mathrm{gal}$ are derived from the galaxy size-magnitude joint distributions measured in the COSMOS field by \cite{Jouvel++09} (and converting from half-light-radius to $r^2$ assuming an exponential galaxy profile), combined with each survey's target number density of galaxies and magnitude limit obtained from \cite{DES++11}, \cite{Chang++13}, and \cite{Laureijs++11}.

\section{Conclusion}\label{sec:conclusion}

Comparing Figures \ref{fig:f1} and \ref{fig:f2} to the rightmost column of in Table \ref{tab:requirements}, we see that the mismatch in PSF size between stars and galaxies from effects such as chromatic seeing for a ground-based telescope and the diffraction limit for a space-based telescope is significantly larger than the requirements for cosmic shear.
Fortunately, corrections can be applied on an object-by-object basis given we have some estimate of each object's SED over the wavelength range of the shape measurement filter(s).
Such estimates are readily available through multifilter photometry.
This approach, roughly akin to photometric redshifts, is studied in more detail in \cite{Cypriano++10, Plazas+Bernstein12, Meyers+Burchat14}.

Corrections of this type require knowledge of the chromatic effects from the entire imaging system, including the atmosphere for ground-based telescopes, and of optics and sensors for both ground and space telescopes.
For the class of chromatic effects investigated here ($\mathrm{FWHM} \propto \lambda^\alpha$), we can estimate how well we need to know the power law index $\alpha$ in order for our correction to succeed.
From Figure \ref{fig:f1} and Table \ref{tab:requirements}, we see that, for LSST $r$-band images, $\delta r^2_\mathrm{PSF}/r^2_\mathrm{PSF}$ varies by a factor of $\sim 25$ more than the requirement.
Since $\delta r^2_\mathrm{PSF}/r^2_\mathrm{PSF}$ varies roughly linearly with $\alpha$, we therefore need to know $\alpha$ to a precision of about $|\alpha|/25 \approx 0.008$.
Similarly, knowledge of $\alpha$ also needs to be at the level of $\approx 0.01$ for Euclid.
Of course, not all chromatic effects are accurately described by a power law model (e.g. \cite{Plazas+Bernstein12}), but this exercise demonstrates the unprecedented requirements for understanding all aspects of future imaging systems.

\acknowledgments
This material is based upon work supported by the National Science Foundation under Grant No. 0969487.

\end{document}